# Prediction of many-electron wavefunctions using atomic potentials: refinements and extensions to transition metals and large systems

Jerry L. Whitten

Department of Chemistry
North Carolina State University
Raleigh, NC 27695 USA

email: whitten@ncsu.edu

**Abstract**

For a given many-electron molecule, it is possible to define a corresponding one-electron Schrödinger equation, using potentials derived from simple atomic densities, whose solution predicts fairly accurate molecular orbitals for single- and multi-determinant wavefunctions for the molecule. The energy is not predicted and must be evaluated by calculating Coulomb and exchange interactions over the predicted orbitals. Transferable potentials for first-row atoms and transition metal oxides that can be used without modification in different molecules are reported. For improved accuracy, molecular wavefunctions can be refined by slightly scaling nuclear charges and by introducing potentials optimized for functional groups. For a test set of 20 molecules representing different bonding environments, the transferable potentials with scaling give wavefunctions with energies that deviate from exact self-consistent field or configuration interaction energies by less than 0.05 eV and 0.02 eV per bond or valence electron pair, respectively. Applications to the ground and excited states of a $Ti_{18}O_{36}$ nanoparticle and chlorophyll-s are reported.

**I. Introduction**

In earlier work, (Nazari and Whitten, 2017), we described a method for predicting wavefunctions using simple atomic potentials.[1] The prediction of wavefunctions by methods other than self-consistent field calculations has a long history and the earlier work describes some of the many ways one can approach the problem such as using hybridized orbitals, localized bond orbitals and expansion methods based on electrostatic error bounds.[2-15] Local bonding constructions can be used to partition large systems into interacting components. Ruedenberg, Head-Gordon and coworkers and have discussed how localized orbitals can be used to construct and analyze molecular wavefunctions.[3-5] Methods have been developed to solve for localized orbitals directly and other investigations have used bonding parameters obtained for molecular fragments or localized components to describe large systems.[2-14] If the objective is simply to determine an initial field for an SCF calculation, there are many options. The simplest is to construct an approximate electron density by summing over atomic densities.[14,15]

In our earlier work and in the present work, a one-particle Schrödinger equation is constructed for a given many-electron system such that its solution matches as closely as possible (to within a unitary transformation) the many-electron SCF solution using the same basis. Simple atomic potentials are shown to exist that predict fairly accurate solutions. These are not densities corresponding to the atomic wavefunction *per se*, but instead are special densities that generate the optimum mixing of basis functions. The existence of such potentials might seem unlikely given the complexity of the Fock operator that correctly determines the mixing of basis functions to form molecular orbitals. However, we show that is not the case, and that one-electron potentials exist that predict remarkably good molecular orbitals for single-determinant and multi-configuration wavefunctions. The energy is not predicted and must be evaluated by including electron repulsion integrals over the molecular orbitals – the result is an upper bound on the exact energy of the system, i.e. the method is variational.

In the present work, we focus on transferable potentials for first-row atoms and transition metal oxides that can be used without modification in different molecules. For improved accuracy, it is shown that molecular wavefunctions can be refined by slightly scaling nuclear charges and by introducing potentials optimized for functional groups such as –COOH and –$NH_2$. The goal is to predict very accurate molecular orbitals for arbitrary systems. For a test set of molecules representing different bonding environments, transferable potentials give wavefunctions with energies that differ from exact SCF or configuration interaction (CI) energies by less than 0.05 eV and 0.02 eV per bond or valence electron pair, respectively. The error is further reduced by scaling nuclear charges. In many cases, the orbitals are found to be accurate enough for direct use in configuration interaction calculations, bypassing completely an SCF calculation. In the present work, potentials are determined in the context of many-electron theory, but the same argument could be applied to density functional calculations.

## II. Method

We begin by considering a molecule or other system described by the Schrödinger equation

$$H_{exact}\psi = (\sum_i [-\tfrac{1}{2}\nabla_i^2 - \sum_q \frac{Z_q}{r_{qi}}] + \sum_{i<j} r_{ij}^{-1})\psi = E_{exact}\psi$$

with electrons and nuclei designated by *i* and *q*, respectively, and associate with the system a modified Hamiltonian, $H^0$, that contains additional one-particle potentials for each nucleus, $v_{qi}$,

$$H^0\psi = (\sum_i h_i)\,\psi = (\sum_i\ [-\tfrac{1}{2}\nabla_i^2 + \sum_q (-\frac{Z_q}{r_{qi}} + v_{qi})])\,\psi = E\psi$$

where $\psi = (norm)\det(\chi_1(1)\chi_2(2)\chi_3(3)...)$ and $\chi_i = \varphi_m(spin)$

Spatial orbitals are obtained by solving the one-electron eigenvalue problem $h_i\varphi_m(i) = \varepsilon_m\varphi_m(i)$.

Two variational solutions, $<\psi|H_{exact}|\psi> \ge E_{exact}$ are of interest for a given basis: a) the single determinant SCF solution and b) a configuration interaction solution involving a selected number of configurations. The objective of the present work is to find potentials $v_{qi}$ that produce orbitals $\varphi_m$ that match as closely possible the variational solutions of a) or b). We first consider the single-determinant solution.

Potentials are assumed to derive from densities centered at nuclei where densities are expanded as a linear combination of normalized spherical Gaussian functions, $\rho_a = (\frac{a}{\pi})^{3/2} \exp(-ar^2)$. For a single component (at nucleus $q$), the repulsive potential acting on particle $i$ is

$$v_i = v(r_i) = \int |\vec{r}_i - \vec{r}|^{-1} \rho_a dv = 2\sqrt{\frac{a}{\pi}} r_i^{-1} \int_0^{r_i} \exp(-ar^2) dr$$

In the present work, up to three such densities are allowed on a given nucleus, $\sum_a c_a \rho_a$. For transition metals, four densities are allowed. It is further assumed that the total density is normalized to the nuclear charge, $Z_q = \sum_a c_a$ such that the nuclear plus the repulsive potential is asymptotically zero for each nucleus. Following the notation of Ref. 1, we refer to the added potential as a QC-potential, Vqc, and the solution of the resulting one-particle Schrödinger equation as the QC-potential method. It might be argued that severe restrictions such as spherical densities and neutral atoms make it unlikely that the resulting potentials would produce accurate results in view of asymmetries in bonding and charge transfer effects. However, spherical densities combined with those on neighboring atoms introduce directional effects, and, varying exponents affects the polarity of bonds, i.e., smaller exponents decrease the shielding making a nucleus more attractive, and conversely. Although the Vqc potentials produce fairly accurate solutions, we have investigated a refinement of the method in which an *in-situ* adjustment of the potential is allowed.

This is accomplished by adding a term to the Vqc potential, $v_i + \frac{\gamma_q}{r_{qi}}$, and optimizing $\gamma_q$ in the molecule. Since this is equivalent to changing the nuclear charge in the one-particle equation, we refer to this as Z-scaling. It should be noted, that the charge neutral transferable potential remains invariant on the addition of this term to $H_0$. There are other ways to allow an *in-situ* variation of the potential such as allowing a change in the exponents or coefficients of the component densities. The proposed variation is one of the simplest and because the departure from neutrality must be small, $\gamma_q$ is small and can be determined very efficiently by direct variation. In systems where there is large charge transfer or where Madelung-like contributions are important, scaling can lead to significant improvement of the predicted wavefunction. After the molecular orbitals are predicted, the QC or Z-scaled potentials no longer contribute since the exact H is used to evaluate the energy.

To reach an acceptable level of accuracy, potentials require careful optimization, but once density parameters have been determined, applications are straightforward. The Nelder-Mead simplex procedure[16] for optimizing densities and corresponding potentials is described in Ref. 1 and is summarized in the Appendix. The result of the optimization is a set of density parameters, $\{c_a, a\}$, for each atom $k$ in the molecule being considered where

$$\rho_q = \sum_a c_a \rho_a = \sum_a c_a (\frac{a}{\pi})^{3/2} \exp(-ar_q^2).$$ To construct average potentials designed to be transferable, we have found it better to consider molecules with complex bonding environments, rather than atoms or simple molecules, so that interactions are averaged over different environments. The present Vqc potentials are based on $C_6H_6$, $N_2C_4H_4$, $H_2NCH_2$-COOH, $C_6H_5$-F, $C_6H_5$-COOH, and HFCO. The first four systems were treated successively, keeping atomic parameters determined for preceding molecules invariant. Parameters for O and F were then averaged with parameters from partial optimizations of the latter two molecules. In all calculations involving potentials, the 1s atomic orbital is constrained to be doubly occupied. This introduces a negligible error when compared to an exact SCF calculation with no 1s occupancy constraint. It is advantageous, however, since the parameters of the QC potential are no longer required to describe core-valence mixing. The 1s-occupancy constraint is the principal reason the present average potentials are slightly more accurate than those in Ref. 1. Lifting the constraint would lead to a slight increase in energy. Densities that generate Vqc potentials are reported in Table 1 along with functional group potentials and an oxygen potential used in Ti, Fe and Ni oxide systems. These potentials are used without modification in all applications.

### III. Basis set

The basis for each atom is a near Hartree Fock set of atomic orbitals plus extra two-component s- and p-type functions consisting of the two smaller exponent components of the Hartree-Fock atomic orbital. We refer to this as a double-zeta basis. The atomic orbitals are expanded as linear combinations of Gaussian functions. The number of components is large since the atomic orbitals are of Hartree-Fock accuracy. The basis can be described as 1s(10), 2s(5), 2p(5), 2s′(2), 2p′(2), for C,N,O , 2p(6) for F, and 1s(4), s(1) for H where the number of Gaussian functions in each orbital is indicated in parentheses. The larger basis set used in the extended treatment of glycine contains d- functions and chlorophyll contains an additional p-orbital in the π-system. For Ti, Fe and Ni, the basis is a highly optimized 1s(12), 2s(10), 2p(7), 3s(7), 3p(6), 3d(4), 3d'(2), 4s(4), 4s'(1), and 4p(2) expansion giving atomic ground states close to the Hartree-Fock limit. The same basis omitting the 4p is used for Ti18O36. In the transition metal systems, the 1s, 2s and 2p atomic orbitals are assigned to an invariant core and all other basis functions are orthogonalized to these orbitals. No core potentials were used in the present calculations so that the predictive capability of the method could be fully tested.

**Table 1.** Densities used to define average potentials for atoms and functional groups Individual densities are normalized, $(\frac{a}{\pi})^{3/2} \exp(-ar^2)$, and coefficients, $c_a$, sum to the nuclear charge except for the Mg-4N functional group. A positive coefficient denotes repulsion.

```
C
exp        9.27799720    0.33295282    0.23450078
coef       2.40426590    3.12627160    0.46946246
H
exp        0.31532871    0.41982534
```

| | | | |
|---|---|---|---|
| coef | 0.99952345 | 0.00047655 | |
| **N** | | | |
| exp | 3.70505490 | 0.35327028 | 0.09079513 |
| coef | 3.67864410 | 2.75698430 | 0.56437154 |
| **O** | | | |
| exp | 5.80696930 | 0.58286375 | 0.11732898 |
| coef | 3.54700310 | 3.50272230 | 0.95027464 |
| **F** | | | |
| exp | 0.87609088 | 19.53952200 | 0.01500000 |
| coef | 7.88516910 | 3.42046720 | 0.69436371 |

**Functional groups**
**Mg-4N(chlorophyll-a)**
**Mg**

| | | | |
|---|---|---|---|
| exp | 1.96671400 | 27.20791200 | 0.14478356 |
| coef | 3.95989600 | 4.93930100 | 4.30345250 |
| **N** | | | |
| exp | 4.64547700 | 0.44293783 | 0.11384088 |
| coef | 3.18098160 | 3.20262660 | 0.21139261 |

**NH (chlorin)**
**N**

| | | | |
|---|---|---|---|
| exp | 16.10528700 | 0.38326927 | 0.04692480 |
| coef | 2.72798330 | 3.98638190 | 0.28563480 |
| **H** | | | |
| exp | 0.50689916 | 0.67488024 | |
| coef | 0.99952345 | 0.00047655 | |

**NH$_2$**
**N**

| | | | |
|---|---|---|---|
| exp | 13.84919100 | 0.74665805 | 0.12246505 |
| coef | 2.16952540 | 3.61395670 | 1.21651790 |
| **H** | | | |
| exp | 0.31532871 | 0.41982534 | |
| coef | 0.99952345 | 0.00047655 | |

**COOH**
**C**

| | | | |
|---|---|---|---|
| exp | 8.33810510 | 0.52182416 | 0.08764625 |
| coef | 1.97136300 | 2.98695760 | 1.04167930 |
| **O** | | | |
| exp | 5.56146050 | 0.21896461 | 0.16644169 |
| coef | 3.96314670 | 12.99709600 | -8.96024290 |
| **OH** | | | |
| exp | 5.33305990 | 0.39949988 | 0.41053428 |
| coef | 3.93729110 | 16.05746800 | -11.99476000 |

**Ti, Fe, Ni oxides**
**O**

|      |              |            |             |             |
|------|-------------:|-----------:|------------:|------------:|
| exp  |   9.00070450 | 0.85820087 |  0.22056896 |             |
| coef |   3.07877500 | 2.42114790 |  2.50007710 |             |
| **Ti** | | | | |
| exp  |  33.42917100 | 0.21901873 |  2.54469520 | 12.12075900 |
| coef |   4.12910470 | 6.99037180 |  7.52713230 |  3.35339120 |
| **Fe** | | | | |
| exp  | 120.61914000 | 0.31700886 |  4.07546880 |  0.31731266 |
| coef |   3.95348930 | 6.47602840 | 14.17785000 |  1.39263260 |
| **Ni** | | | | |
| exp  | 158.46162000 | 0.14067395 |  7.07254620 |  0.83915098 |
| coef |   3.96417380 | 2.77357140 | 11.97728600 |  9.28496920 |

## IV. Applications

In this section, applications using the Vqc densities (potentials) defined in Table I are reported for a test set of 20 molecules. Refinements using functional group potentials and *in situ* Z-scaling of the nuclear charge are also included. Wavefunctions are predicted, energies are evaluated, and results are compared with all-electron SCF and CI calculations using the same basis.[17] We refer to the latter canonical SCF solution and CI calculation as "exact" for the given basis. All atoms with the same atomic number in the same molecule have the same Vqc potential. This constraint is relaxed when functional group potentials are used, and the local environment is further improved by Z-scaling. Since the exact Hamiltonian is used to evaluate the energy of the predicted wavefunction, the energy is variational.

In the CI calculations, the SCF step is eliminated completely and molecular orbitals produced by solving the one-electron potential problem are used directly in the CI. As noted earlier, the QC-method only predicts orbitals: Coulomb and exchange integrals over molecular orbitals are required for the single-determinant energy and all two-electron integrals are needed for the CI calculations. The "exact" CI calculations used for comparison are performed using canonical SCF occupied and virtual orbitals, i.e., no transformation of orbitals is used to improve the convergence.

The plan is as follows: First, calculations on several representative molecules are discussed in detail beginning with the use of Vqc potentials and then including functional group and Z-scaling refinements. The results of these studies are reported in Table 2. Results for the larger test set of molecules are summarized in Table 3. Calculations on NiO, FeO and TiO$_2$ are then discussed and an application of the Ti and oxide potential to an all-electron treatment of the ground and excited states of Ti$_{18}$O$_{36}$ using a flexible double-zeta set of d- orbitals is described. The present study concludes with a treatment of the ground and excited states of chlorophyll-a using functional group Vqc potentials and Z-scaling.

**Calculations on representative molecules**

In this section, calculations on several representative molecules are described. The hierarchy in accuracy of the calculations and entries in Table 2 are as follows:
1) transfer potentials (from Table 1)
2) functional group potentials (from Table 1)
3) Z-scaling by variation of Z

**Table 2.** Comparison of energies for wavefunctions predicted by atomic and group QC-potentials including Z scaling with exact SCF and CI calculations for selected molecules. [a]

| | total E[b] | | | error[b] | | | error (%) | | | error[c] |
|---|---|---|---|---|---|---|---|---|---|---|
| molecule/method | 1-det | small CI | large CI | 1-det | small CI | large CI | 1-det | small CI | large CI | eV/val pair |
| **$H_2CO$** | | | | | | | | | | |
| exact | -113.8287 | -113.9980 | -114.0575 | | | | | | | |
| Vqc | -113.8196 | -113.9952 | -114.0580 | -0.0091 | -0.0027 | 0.0005 | 0.0080 | 0.0024 | 0.0 | 0.0 |
| Vqc + Z | -113.8231 | -113.9969 | -114.0572 | -0.0056 | -0.0011 | -0.0003 | 0.0050 | 0.0009 | 0.0002 | 0.0012 |
| **$C_5H_5N$** | | | | | | | | | | |
| exact | -246.6176 | -246.7239 | -247.1093 | | | | | | | |
| Vqc | -246.6011 | -246.7232 | -247.1033 | -0.0165 | -0.0007 | -0.0059 | 0.0067 | 0.0003 | 0.0024 | 0.0108 |
| Vqc + Z | -246.6053 | -246.7167 | -247.1110 | -0.0122 | -0.0072 | 0.0017 | 0.0050 | 0.0029 | 0.0 | 0.0 |
| **$NH_2$-$CH_2$-COOH (glycine)** | | | | | | | | | | |
| exact | -282.7387 | -282.7692 | -283.2125 | | | | | | | |
| Vqc | -282.7068 | -282.7485 | -283.2019 | -0.0319 | -0.0207 | -0.0106 | 0.0113 | 0.0073 | 0.0037 | 0.0192 |
| opt $NH_2$, COOH | -282.7156 | -282.7577 | -283.2033 | -0.0231 | -0.0116 | -0.0093 | 0.0082 | 0.0041 | 0.0033 | 0.0168 |
| Vqc + Z | -282.7183 | -282.7615 | -283.2083 | -0.0204 | -0.0077 | -0.0042 | 0.0072 | 0.0027 | 0.0015 | 0.0077 |
| Vgroup+ Z | -282.7214 | -282.7600 | -283.2071 | -0.0173 | -0.0093 | -0.0055 | 0.0061 | 0.0033 | 0.0019 | 0.0100 |
| **$C_6H_5$-COOH** | | | | | | | | | | |
| exact | -418.1783 | -418.2621 | -418.8724 | | | | | | | |
| Vqc | -418.1341 | -418.2503 | -418.8567 | -0.0442 | -0.0118 | -0.0156 | 0.0106 | 0.0028 | 0.0037 | 0.0185 |
| opt COOH | -418.1531 | -418.2467 | -418.8589 | -0.0252 | -0.0155 | -0.0134 | 0.0060 | 0.0037 | 0.0032 | 0.0159 |
| Vqc + Z | -418.1523 | -418.2378 | -418.8569 | -0.0260 | -0.0243 | -0.0155 | 0.0062 | 0.0058 | 0.0037 | 0.0184 |
| Vgroup+ Z | -418.1571 | -418.2492 | -418.8604 | -0.0212 | -0.0129 | -0.0120 | 0.0051 | 0.0031 | 0.0029 | 0.0142 |
| **$C_2F_2H_2$** | | | | | | | | | | |
| exact | -275.6546 | -275.8525 | -276.0854 | | | | | | | |
| Vqc | -275.6221 | -275.8623 | -276.0801 | -0.0326 | 0.0098 | -0.0052 | 0.0118 | 0.0 | 0.0019 | 0.0119 |
| Vqc + Z | -275.6381 | -275.8478 | -276.0817 | -0.0166 | -0.0047 | -0.0036 | 0.0060 | 0.0017 | 0.0013 | 0.0082 |
| **$N_4C_{20}H_{16}$ (chlorin)** | | | | | | | | | | |
| exact | -984.2629 | -984.2959 | -984.5745 | | | | | | | |
| Vqc | -984.1589 | -984.2323 | -984.5295 | -0.1041 | -0.0636 | -0.0451 | 0.0106 | 0.0065 | 0.0046 | 0.0211 |
| opt NH | -984.1795 | -984.2370 | -984.5361 | -0.0834 | -0.0589 | -0.0384 | 0.0085 | 0.0060 | 0.0039 | 0.0180 |
| Vqc + Z | -984.1849 | -984.2361 | -984.5498 | -0.0781 | -0.0598 | -0.0247 | 0.0079 | 0.0061 | 0.0025 | 0.0116 |
| Vgroup+ Z | -984.1851 | -984.2377 | -984.5483 | -0.0778 | -0.0582 | -0.0262 | 0.0079 | 0.0059 | 0.0027 | 0.0123 |

[a] The Vqc potential for an atom is the same in each molecule as defined in Table 1.

[b] Energies are in hartrees. A negative value for the error indicates that the exact value is the lowest energy.

[c] Errors per valence electron pair are calculated from the large CI energies.

Applications 1) and 2) are simple since only one single-particle equation is solved. Calculation 3) requires repetition of the single-particle solution in the molecule as charges are varied. The latter calculation can be done quickly by various numerical methods (the simplex method is used in the present work) since the variations in charge are small, ~ 0.01. Note that only the energy requires evaluation in the many-electron calculation and no Fock matrix construction is necessary. Important conclusions from the results in Table 2 can be summarized as follows:

1) The 1-determinant energies from the Vqc potentials differ from the exact by less than 0.012%, 100(error/total E); for those molecules with a functional group potential, the error is reduced to 0.0085%. Error limits are approximately the same for the full test set of molecules as reported in Table 3. These 1-determinant errors translate into a maximum error of 0.05 eV per valence electron pair for all molecules studied. If the purpose of the calculation is only to generate an initial field for an SCF calculation, then the Vqc method provides an excellent solution.

2) The atomic and functional group potentials show remarkably small errors at the small CI level (~$10^2$ dets) differing from the exact by only 0.007% . For the large CI ( ~$10^4$ dets) the maximum error of 0.005% occurs for chlorin. Similar CI errors are also found for the larger test set of molecules. The worst case corresponds to an error of only 0.02 eV/valence electron pair.

3) For the smaller molecules in Table 2, there is no error at the CI level and the Vqc orbitals may give a lower energy than obtained for the canonical SCF orbitals. Canonical SCF orbitals produce virtual orbitals that are equivalent to negative ion orbitals and are often too diffuse spatially for optimum correlation by CI expansions. This is a well-known effect and virtual orbitals are often transformed by positive-ion calculations or exchange maximization to improve convergence. The Vqc orbitals, since they arise from a one-particle potential, are the same "quality" for occupied and unoccupied orbitals. Thus, the energy of the leading determinant is slightly higher for the Vqc solution, but the CI convergence is improved compared to that based on canonical SCF orbitals.

4) The errors for the large CI calculations are small and to the extent that such errors are tolerable, this means that the Vqc orbitals can be used directly in the CI or for other orbital based correlation methods eliminating the SCF step completely.

5) The results in both tables show Z-scaling can significantly reduce the error of the single-determinant solution. For all molecules studied, the CI energy is also improved, but often only slightly. The Z-scaling refinement becomes more worthwhile if there are unusual bonding environments such as ionic charge distributions or if the correlation method requires a better leading determinant solution.

Of course, it is not surprising that the CI errors are greatly reduced compared to the single-determinant errors since CI expansions recover part of the defect in molecular orbitals. If the CI calculations were complete all molecular orbital sets that are related by a linear transformation would give the same result. In general, CI expansions are not complete, however. In the present calculations, configurations are generated by a hierarchical procedure[17] that includes excitations from determinants in the expansion with coefficients greater than 0.02 if the second order energy of interaction with the initial expansion exceeds $1 \times 10^{-6}$ hartrees. The resulting expansions contain $10^4$ - $10^5$ determinants in the test set of molecules and thus the treatments are not near the full CI limit. The fact that small residual errors exist means that defects in occupied orbitals are not fully recovered. For the larger systems of chlorin, $Ti_{18}O_{36}$ and chlorophyll, the latter two will be discussed later, the active CI space does not include all electrons of the system and adjustments of the lower energy orbitals cannot occur. Thus, it is encouraging that the CI errors turn out to be relatively small for these systems.

**Table 3.** Comparison of energies for wavefunctions predicted by atomic and group QC-potentials including Z scaling with exact SCF and CI calculations for all molecules studied excluding transition metals. [a]

| molecule/method | total E[b] | | | error[b] | | | error(%) | | | error[c] |
|---|---|---|---|---|---|---|---|---|---|---|
| | 1-det | small CI | large CI | 1-det | small CI | large CI | 1-det | small CI | large CI | eV/val pair |
| **C$_6$H$_6$** | | | | | | | | | | |
| exact | -230.6485 | -230.7559 | -231.1362 | | | | | | | |
| Vqc | -230.6397 | -230.7579 | -231.1340 | -0.0088 | 0.0020 | -0.0022 | 0.0038 | 0.0 | 0.0009 | 0.0039 |
| Vqc + Z | -230.6429 | -230.7513 | -231.1349 | -0.0056 | -0.0046 | -0.0012 | 0.0024 | 0.0020 | 0.0005 | 0.0022 |
| **C$_4$H$_4$N$_2$** | | | | | | | | | | |
| exact | -262.5793 | -262.7274 | -263.0907 | | | | | | | |
| Vqc | -262.5563 | -262.7303 | -263.0844 | -0.0230 | 0.0029 | -0.0063 | 0.0088 | 0.0 | 0.0024 | 0.0114 |
| Vqc + Z | -262.5626 | -262.7235 | -263.0859 | -0.0167 | -0.0038 | -0.0048 | 0.0064 | 0.0015 | 0.0018 | 0.0087 |
| **C$_5$H$_5$N** | | | | | | | | | | |
| exact | -246.6176 | -246.7239 | -247.1093 | | | | | | | |
| Vqc | -246.6011 | -246.7232 | -247.1033 | -0.0165 | -0.0007 | -0.0059 | 0.0067 | 0.0003 | 0.0024 | 0.0108 |
| Vqc + Z | -246.6053 | -246.7167 | -247.1110 | -0.0122 | -0.0072 | 0.0017 | 0.0050 | 0.0029 | 0.0 | 0.0 |
| **C$_2$H$_4$** | | | | | | | | | | |
| exact | -78.0194 | -78.1477 | -78.2247 | | | | | | | |
| Vqc | -78.0181 | -78.1494 | -78.2246 | -0.0013 | 0.0016 | -0.0001 | 0.0017 | 0.0 | 0.0001 | 0.0005 |
| Vqc + Z | -78.0181 | -78.1472 | -78.2247 | -0.0013 | -0.0005 | 0.0000 | 0.0016 | 0.0007 | 0.0 | 0.0001 |
| **CH$_4$** | | | | | | | | | | |
| exact | -40.1874 | -40.2767 | -40.3032 | | | | | | | |
| Vqc | -40.1800 | -40.2730 | -40.3043 | -0.0073 | -0.0037 | 0.0011 | 0.0183 | 0.0092 | 0.0 | 0.0 |
| Vqc + Z | -40.1871 | -40.2770 | -40.3037 | -0.0003 | 0.0003 | 0.0005 | 0.0008 | 0.0 | 0.0 | 0.0 |
| **C$_2$H$_2$** | | | | | | | | | | |
| exact | -76.8089 | -76.9609 | -77.0112 | | | | | | | |
| Vqc | -76.8050 | -76.9520 | -77.0109 | -0.0039 | -0.0089 | -0.0003 | 0.0051 | 0.0116 | 0.0004 | 0.0017 |
| Vqc + Z | -76.8066 | -76.9600 | -77.0113 | -0.0023 | -0.0009 | 0.0001 | 0.0030 | 0.0012 | 0.0 | 0.0 |

**$H_2O$**

| | | | | | | | | | | |
|---|---:|---:|---:|---:|---:|---:|---:|---:|---:|---:|
| exact | -76.0079 | -76.1213 | -76.1379 | | | | | | | |
| Vqc | -76.0006 | -76.1215 | -76.1382 | -0.0073 | 0.0002 | 0.0003 | 0.0096 | 0.0 | 0.0 | 0.0 |
| Vqc + Z | -76.0057 | -76.1189 | -76.1379 | -0.0022 | -0.0024 | 0.0000 | 0.0028 | 0.0032 | 0.0 | 0.0001 |

**$H_2CO$**

| | | | | | | | | | | |
|---|---:|---:|---:|---:|---:|---:|---:|---:|---:|---:|
| exact | -113.8287 | -113.9980 | -114.0575 | | | | | | | |
| Vqc | -113.8196 | -113.9952 | -114.0580 | -0.0091 | -0.0027 | 0.0005 | 0.0080 | 0.0024 | 0.0 | 0.0 |
| Vqc + Z | -113.8231 | -113.9969 | -114.0572 | -0.0056 | -0.0011 | -0.0003 | 0.0050 | 0.0009 | 0.0002 | 0.0012 |

**CO**

| | | | | | | | | | | |
|---|---:|---:|---:|---:|---:|---:|---:|---:|---:|---:|
| exact | -112.6983 | -112.8889 | -112.8941 | | | | | | | |
| Vqc | -112.6822 | -112.8914 | -112.8989 | -0.0162 | 0.0025 | 0.0048 | 0.0144 | 0.0 | 0.0 | 0.0 |
| Vqc + Z | -112.6924 | -112.8918 | -112.8981 | -0.0059 | 0.0029 | 0.0040 | 0.0053 | 0.0 | 0.0 | 0.0 |

**$NC_4H_5$**

| | | | | | | | | | | |
|---|---:|---:|---:|---:|---:|---:|---:|---:|---:|---:|
| exact | -208.7742 | -208.8384 | -209.1913 | | | | | | | |
| Vqc | -208.7562 | -208.8499 | -209.1844 | -0.0180 | 0.0114 | -0.0068 | 0.0086 | 0.0 | 0.0033 | 0.0143 |
| Vqc + Z | -208.7644 | -208.8340 | -209.1877 | -0.0098 | -0.0045 | -0.0035 | 0.0047 | 0.0021 | 0.0017 | 0.0074 |

**$NC_4H_4$**

| | | | | | | | | | | |
|---|---:|---:|---:|---:|---:|---:|---:|---:|---:|---:|
| exact | -208.1265 | -208.1787 | -208.5347 | | | | | | | |
| Vqc | -208.0946 | -208.1684 | -208.5276 | -0.0320 | -0.0103 | -0.0071 | 0.0154 | 0.0050 | 0.0034 | 0.0155 |
| Vqc + Z | -208.1079 | -208.1822 | -208.5296 | -0.0186 | 0.0035 | -0.0051 | 0.0089 | 0.0 | 0.0025 | 0.0112 |

**$NH_2$-$CH_2$-COOH (glycine)**

| | | | | | | | | | | |
|---|---:|---:|---:|---:|---:|---:|---:|---:|---:|---:|
| exact | -282.7387 | -282.7692 | -283.2125 | | | | | | | |
| Vqc | -282.7068 | -282.7485 | -283.2019 | -0.0319 | -0.0207 | -0.0106 | 0.0113 | 0.0073 | 0.0037 | 0.0192 |
| opt $NH_2$, COOH | -282.7156 | -282.7577 | -283.2033 | -0.0231 | -0.0116 | -0.0093 | 0.0082 | 0.0041 | 0.0033 | 0.0168 |
| Vqc + Z | -282.7183 | -282.7615 | -283.2083 | -0.0204 | -0.0077 | -0.0042 | 0.0072 | 0.0027 | 0.0015 | 0.0077 |
| | Vgroup+ Z | -282.7214 | -282.7600 | -283.2071 | -0.0173 | -0.0093 | -0.0055 | 0.0061 | 0.0033 | 0.0019 | 0.0100 |

**C$_6$H$_5$-NH$_2$**

| | | | | | | | | | | |
|---|---|---|---|---|---|---|---|---|---|---|
| exact | -285.6597 | -285.7295 | -286.2190 | | | | | | | |
| Vqc | -285.6442 | -285.7142 | -286.2110 | -0.0156 | -0.0154 | -0.0080 | 0.0054 | 0.0054 | 0.0028 | 0.0121 |
| opt NH$_2$ | -285.6465 | -285.7119 | -286.2132 | -0.0132 | -0.0176 | -0.0058 | 0.0046 | 0.0062 | 0.0020 | 0.0088 |
| Vqc + Z | -285.6483 | -285.7224 | -286.2157 | -0.0114 | -0.0071 | -0.0033 | 0.0040 | 0.0025 | 0.0012 | 0.0050 |
| Vgroup+ Z | -285.6487 | -285.7171 | -286.2126 | -0.0110 | -0.0125 | -0.0065 | 0.0039 | 0.0044 | 0.0023 | 0.0098 |

**C$_5$H$_5$-COOH**

| | | | | | | | | | | |
|---|---|---|---|---|---|---|---|---|---|---|
| exact | -380.2957 | -380.3619 | -380.9477 | | | | | | | |
| Vqc | -380.2585 | -380.3284 | -380.9271 | -0.0372 | -0.0335 | -0.0205 | 0.0098 | 0.0088 | 0.0054 | 0.0279 |
| opt COOH | -380.2728 | -380.3479 | -380.9367 | -0.0228 | -0.0140 | -0.0109 | 0.0060 | 0.0037 | 0.0029 | 0.0149 |
| Vqc + Z | -380.2727 | -380.3506 | -380.9376 | -0.0229 | -0.0113 | -0.0101 | 0.0060 | 0.0030 | 0.0026 | 0.0137 |
| Vgroup+ Z | -380.2759 | -380.3544 | -380.9377 | -0.0198 | -0.0075 | -0.0100 | 0.0052 | 0.0020 | 0.0026 | 0.0136 |

**C$_6$H$_5$-COOH**

| | | | | | | | | | | |
|---|---|---|---|---|---|---|---|---|---|---|
| exact | -418.1783 | -418.2621 | -418.8724 | | | | | | | |
| Vqc | -418.1341 | -418.2503 | -418.8567 | -0.0442 | -0.0118 | -0.0156 | 0.0106 | 0.0028 | 0.0037 | 0.0185 |
| opt COOH | -418.1531 | -418.2467 | -418.8589 | -0.0252 | -0.0155 | -0.0134 | 0.0060 | 0.0037 | 0.0032 | 0.0159 |
| Vqc + Z | -418.1523 | -418.2378 | -418.8569 | -0.0260 | -0.0243 | -0.0155 | 0.0062 | 0.0058 | 0.0037 | 0.0184 |
| Vgroup+ Z | -418.1571 | -418.2492 | -418.8604 | -0.0212 | -0.0129 | -0.0120 | 0.0051 | 0.0031 | 0.0029 | 0.0142 |

**FHCO**

| | | | | | | | | | | |
|---|---|---|---|---|---|---|---|---|---|---|
| exact | -212.6782 | -212.8580 | -213.0189 | | | | | | | |
| Vqc | -212.6544 | -212.8651 | -213.0200 | -0.0238 | 0.0070 | 0.0010 | 0.0112 | 0.0 | 0.0 | 0.0 |
| Vqc + Z | -212.6617 | -212.8593 | -213.0179 | -0.0164 | 0.0013 | -0.0010 | 0.0077 | 0.0 | 0.0 | 0.0030 |

**C$_2$F$_2$H$_2$**

| | | | | | | | | | | |
|---|---|---|---|---|---|---|---|---|---|---|
| exact | -275.6546 | -275.8525 | -276.0854 | | | | | | | |
| Vqc | -275.6221 | -275.8623 | -276.0801 | -0.0326 | 0.0098 | -0.0052 | 0.0118 | 0.0 | 0.0019 | 0.0119 |
| Vqc + Z | -275.6381 | -275.8478 | -276.0817 | -0.0166 | -0.0047 | -0.0036 | 0.0060 | 0.0017 | 0.0013 | 0.0082 |

**C₆H₅-F**

| | | | | | | | | | |
|---|---:|---:|---:|---:|---:|---:|---:|---:|---:|
| exact | -329.5020 | -329.6147 | -330.0797 | | | | | | |
| Vqc | -329.4758 | -329.5823 | -330.0697 | -0.0262 | -0.0324 | -0.0100 | 0.0079 | 0.0098 | 0.0030 | 0.0151 |
| Vqc + Z | -329.4867 | -329.6017 | -330.0736 | -0.0153 | -0.0130 | -0.0060 | 0.0047 | 0.0039 | 0.0018 | 0.0091 |

**C₂₄H₁₂ (graphene model)**

| | | | | | | | | | |
|---|---:|---:|---:|---:|---:|---:|---:|---:|---:|
| exact | -915.6445 | -915.7236 | -916.0835 | | | | | | |
| Vqc | -915.6011 | -915.6944 | -916.0709 | -0.0434 | -0.0292 | -0.0126 | 0.0047 | 0.0032 | 0.0014 | 0.0064 |
| Vqc + Z | -915.6129 | -915.6958 | -916.0679 | -0.0316 | -0.0278 | -0.0156 | 0.0035 | 0.0030 | 0.0017 | 0.0079 |

**N₄C₂₀H₁₆ (chlorin)**

| | | | | | | | | | |
|---|---:|---:|---:|---:|---:|---:|---:|---:|---:|
| exact | -984.2629 | -984.2959 | -984.5745 | | | | | | |
| Vqc | -984.1589 | -984.2323 | -984.5295 | -0.1041 | -0.0636 | -0.0451 | 0.0106 | 0.0065 | 0.0046 | 0.0211 |
| opt NH | -984.1795 | -984.2370 | -984.5361 | -0.0834 | -0.0589 | -0.0384 | 0.0085 | 0.0060 | 0.0039 | 0.0180 |
| Vqc + Z | -984.1849 | -984.2361 | -984.5498 | -0.0781 | -0.0598 | -0.0247 | 0.0079 | 0.0061 | 0.0025 | 0.0116 |
| Vgroup+ Z | -984.1851 | -984.2377 | -984.5483 | -0.0778 | -0.0582 | -0.0262 | 0.0079 | 0.0059 | 0.0027 | 0.0123 |

**C₄H₄N₂**   incl d

| | | | | | | | | | |
|---|---:|---:|---:|---:|---:|---:|---:|---:|---:|
| exact | -262.6582 | -262.7680 | -263.2504 | | | | | | |
| Vqc | -262.6247 | -262.7645 | -263.2372 | -0.0335 | -0.0035 | -0.0132 | 0.0128 | 0.0013 | 0.0050 | 0.0240 |
| Vqc + Z | -262.6322 | -262.7659 | -263.2408 | -0.0260 | -0.0021 | -0.0096 | 0.0099 | 0.0008 | 0.0036 | 0.0170 |

**NH₂-CH₂-COOH (glycine)**   3-zeta

| | | | | | | | | | |
|---|---:|---:|---:|---:|---:|---:|---:|---:|---:|
| exact | -282.7876 | -282.8051 | -283.2629 | | | | | | |
| Vqc | -282.7427 | -282.7803 | -283.2465 | -0.0450 | -0.0248 | -0.0227 | 0.0159 | 0.0088 | 0.0080 | 0.0410 |
| opt NH₂, COOH | -282.7560 | -282.7844 | -283.2532 | -0.0316 | -0.0206 | -0.0160 | 0.0112 | 0.0073 | 0.0056 | 0.0290 |
| Vqc + Z | -282.7586 | -282.7951 | -283.2564 | -0.0290 | -0.0099 | -0.0128 | 0.0103 | 0.0035 | 0.0045 | 0.0230 |
| Vgroup+ Z | -282.7634 | -282.7823 | -283.2571 | -0.0242 | -0.0227 | -0.0121 | 0.0086 | 0.0080 | 0.0043 | 0.0220 |

[a] The Vqc potential for a given atom is the same in each molecule as defined in Table 1.
[b] Energies are in hartrees. A negative value for the error indicates that the exact value is the lowest energy.
[c] Errors per valence electron pair are calculated from the large CI energies.

**Transition metal oxides**

Transition metal systems are more difficult to treat by electronic structure methods because of closely spaced states arising from different d-orbital occupancies and spin multiplicities. An SCF calculation on the lowest energy state produces orbitals that are not optimum for a different d-occupancy and to describe multiple states may require many determinants to recover from this defect. In this section, we consider several transition metal oxides and describe all systems using the oxygen (oxide) potential defined in Table 1. The Vqc potentials for Ni, Fe and Ti were obtained by optimizing the triplet ground states of NiO, FeO, TiO keeping the oxide potential invariant. The resulting potentials are also given in Table 1. These potentials were then used without change to describe ground and excited states of NiO, FeO, TiO2 and the nanoparticle $Ti_{18}O_{36}$. Related work on metal-metal bonding for transition metals is in progress.[18]

The oxide calculations are summarized in Table 4. Several points are noteworthy. First, a Vqc potential may not automatically produce the lowest energy state if the molecular orbitals are occupied in the order produced by the single particle Hamiltonian. This is not an obstacle, since the occupancy can be selected by inspection or obtained by performing a small CI calculation. Both single-determinant and small CI energies are reported in Table 4 for the various states. Second, for all systems studied, the Vqc molecular orbitals provide a better choice of orbitals for the CI expansion than those of the canonical SCF solution. As pointed out earlier, the latter virtual orbitals correspond to negative ion orbitals and often are too diffuse spatially to provide a good expansion basis. In the limit of a full CI calculation, none of these details matter since the resulting energies and wavefunctions are the same for any set of molecular orbitals related by a linear transformation. However, in practice, when the CI calculation is incomplete, an improved set of molecular orbitals is advantageous. When a Vqc potential is used, the leading determinant has a slightly higher energy than for a canonical SCF calculation, but the CI convergence is faster. Note that even at the small CI level of a few hundred determinants, the Vqc energies are comparable to or better than the canonical SCF values.

**Table 4.** Ground and excited states of several transition metal oxides. Energies from CI calculations using molecular orbitals from The Vqc potentials are compared with values obtained from canonical SCF orbitals.[a] There are many other low lying singlet and higher spin states in the small molecules and only triplet states are listed; singlet states are included for the nanoparticle.

| | | exact[a] | | | Vqc | | | error[b] | |
|---|---|---|---|---|---|---|---|---|---|
| | state | 1-det | small CI ~ $10^2$ dets | large CI ~ $10^5$ dets | 1-det | small CI | large CI | small CI ~ $10^2$ dets | large CI ~ $10^5$ dets |
| **NiO** | 1 | -1581.6123 | -1581.7903 | -1581.9014 | -1581.4408 | -1581.7545 | -1581.9037 | -0.0358 | 0.0024 |
| | 2 | -1581.5961 | -1581.7871 | -1581.8963 | -1581.4225 | -1581.7355 | -1581.8917 | -0.0516 | -0.0046 |
| | 3 | -1581.0877 | -1581.7581 | -1581.8834 | -1581.3604 | -1581.7206 | -1581.8838 | -0.0375 | 0.0004 |
| **FeO** | 1 | -1337.0062 | -1337.2030 | -1337.3734 | -1336.9767 | -1337.2057 | -1337.3786 | 0.0027 | 0.0051 |
| | 2 | -1336.8490 | -1337.1530 | -1337.3268 | -1336.9768 | -1337.1804 | -1337.3366[c] | 0.0274 | 0.0098 |
| | 3 | -1336.8136 | -1337.0850 | -1337.2653 | -1336.9658 | -1337.0750 | -1337.3025 | -0.0100 | 0.0371 |
| **TiO₂** | 1 | -998.0069 | -998.1526 | -998.3856 | -997.8681 | -998.1457 | -998.3905 | -0.0069 | 0.0049 |
| | 2 | -997.7103 | -998.1232 | -998.3631 | -997.9287 | -998.1159 | -998.3659 | -0.0073 | 0.0028 |
| | 3 | -997.8030 | -998.1185 | -998.3398 | -997.7175 | -998.0857 | -998.3493 | -0.0328 | 0.0095 |

| | | | | | | | | transition energies (eV) | |
|---|---|---|---|---|---|---|---|---|---|
| **Ti₁₈O₃₆** | | | | | | | | exact | Vqc |
| | singlet states | | | | | | | | |
| | 1 | -17967.3471 | -17967.3471 | -17967.3575 | -17966.8723 | -17966.8870 | -17966.9207 | | |
| | 2 | -17967.0944 | -17967.1406 | -17967.1770 | -17966.6172 | -17966.6897 | -17966.7511 | 4.91 | 4.62 |
| | 3 | -17967.0831 | -17967.1393 | -17967.1654 | -17966.5938 | -17966.6808 | -17966.7388 | 5.23 | 4.95 |
| | triplet states | | | | | | | | |
| | 1 | -17967.0956 | -17967.1475 | -17967.1797 | -17966.6177 | -17966.6957 | -17966.7546 | 4.84 | 4.52 |
| | 2 | -17967.0844 | -17967.1370 | -17967.1787 | -17966.5947 | -17966.6856 | -17966.7455 | 4.86 | 4.77 |

[a]Exact indicates a calculation based on molecular orbitals from a canonical SCF calculation.

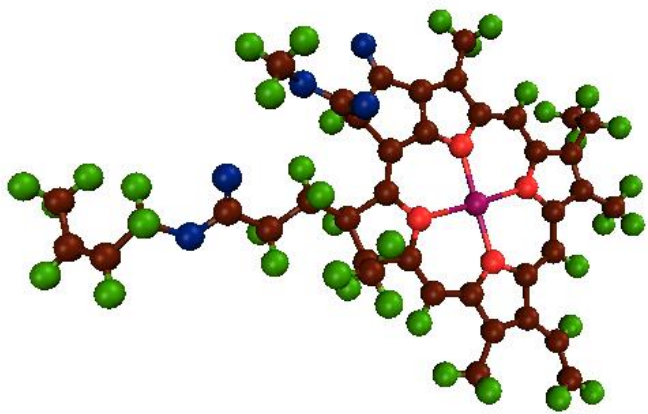

**Figure 1.** Chlorophyll-a. A group potential is used for the central Mg and four nitrogen subsystem.

### Chlorophyll-a

In this section, we consider the ground and low-lying singlet and triplet states of chlorophyll-a. The Mg-4N- region of the molecule, depicted in Fig. 1 is a major factor in determining the properties of the molecule. Instead of developing a potential for Mg, we determine instead a potential for the Mg-4N functional group by direct optimization of the chlorophyll molecule in which Vqc potentials for C, O, and H are used for the remainder of the system. The resulting functional group potential is used along with the other potentials to calculate the ground state wavefunction (molecular orbitals) of the molecule. The energies of the ground and excited state are then calculated using these molecular orbitals and the exact Hamiltonian, i.e., calculating all electron repulsion integrals over molecular orbitals. Because of the large size of the system, of the 185 doubly occupied orbitals in the ground state, 167 orbitals are taken to be in an invariant core and only a total of 99 occupied and virtual orbitals are used to describe the 36-electron active space in the CI calculations. Table 5 summarizes the calculations on the ground and excited states using the invariant potentials. First, we note that the ground state total energy differs by 0.317 hartrees from the exact single determinant SCF calculation, an error of 0.013 %, comparable to that found for the molecules discussed previously. The transition energies to excited states reported in the Table 5 are in good agreement with the values from canonical SCF orbitals for all states considered.

Next, we refine the description of the molecule by Z-scaling of selected atoms, i.e., Mg, N, and the three types of O atoms that occur in the molecule (a total of eight $\frac{\gamma_q}{r_{qi}}$ potentials). The ground and excited states from this calculation are also reported in Table 5. Although the single determinant energy improves slightly, there is only a minor change in transition energies. Thus, the Vqc and Z-scaling treatments and the exact calculation are all in fairly good agreement. As noted earlier, the somewhat smaller excitation energies from the Vqc and scaling treatments may be more accurate than those obtained using canonical SCF orbitals from the ground state SCF solution.

**Table 5.** Ground and excited states of chlorophyll-a. Energies from 1-det and CI calculations using molecular orbitals from Vqc potentials including Z-scaling are compared with values obtained from canonical SCF orbitals. A group potential is used for Mg-4N, as defined in Table 1.

|  | total energy[a] | | | excitation energies (eV) | | |
| --- | --- | --- | --- | --- | --- | --- |
|  | exact | Mg-4N | Mg-4N vary Z | exact | Mg-4N | Mg-4N vary Z |
| **1-det** | -2368.9194 | -2368.6024 | -2368.6420 |  |  |  |
| singlet gnd | -2369.1419 | -2368.8840 | -2368.9149 |  |  |  |
| triplet | -2369.0655 | -2368.8109 | -2368.8432 | 2.08 | 1.99 | 1.95 |
| triplet | -2369.0520 | -2368.7997 | -2368.8342 | 2.45 | 2.29 | 2.20 |
| singlet | -2369.0333 | -2368.7730 | -2368.8072 | 2.95 | 3.02 | 2.93 |
| triplet | -2369.0287 | -2368.7607 | -2368.8002 | 3.08 | 3.35 | 3.12 |
| singlet | -2369.0244 | -2368.7592 | -2368.8001 | 3.20 | 3.40 | 3.12 |

[a] Total energies are in hartrees.

## V. Conclusions

For every molecule, we postulate the existence of a one-electron Schrödinger equation, defined by special (Vqc) potentials, the solution of which generates a useful wavefunction (molecular orbitals) for the many-electron molecule. These special potentials are reported for individual atoms and functional groups. The potentials can be transferred to different molecules without modification. For a test set of molecules, including the large molecule chlorophyll-a, the energies of the predicted wavefunctions differ from the exact values by a maximum of 0.05 eV/valence electron pair and 0.02 eV/valence electron pair for single-determinant and CI wavefunctions, respectively. The molecular orbitals of the wavefunction are predicted by the potential method, but the energy calculations require all electron repulsion integrals. If an improved single-determinant wavefunction is desired, it is possible to refine the description by *in-situ* variation of additional $\gamma/r$ potentials in a molecule. Because the values of the $\gamma$ charges are small, this calculation can be performed efficiently. Results of this Z-scaling refinement are reported for all molecules including the large systems chlorophyll-a and the $Ti_{18}O_{36}$ nanoparticle. Potentials for transition metals in several transition metal oxides are reported along with calculations on several low-lying excited states. The accuracy the Vqc method for the transition metal systems is comparable to that for the other systems. If one wishes to carry out a many-electron SCF calculation, the results of the potential method provide an excellent initial field. If a one-electron Hamiltonian is needed to describe a portion of a large system by other matrix element methods, the Vqc potentials provide a way to determine matrix elements. In future work, the potentials reported should be applied to a larger data base of molecules and their utility explored within a DFT framework.

**Appendix**

Procedure for optimizing densities:

1) Initial exponents and coefficients are specified as parameters for the constituent atomic densities (e.g., for benzene, exponents and coefficients for C and H densities). Suppose the parameter values are $w_1, w_2, w_3 \ldots w_p$.

2) The resulting one-electron eigenvalue problem is solved to determine energies and coefficients of basis functions in molecular orbitals, $\{\varepsilon_m, \varphi_m\}$. The lowest energy N spin orbitals are occupied, (e.g., for benzene, 21 spatial molecular orbitals).

3) A single determinant wavefunction is constructed from the predicted orbitals and its energy is evaluated using the exact Hamiltonian, $H_{exact}$, a step that requires all electron repulsion integrals. The energy, $<\psi | H_{exact} | \psi>$, is a function of the parameters, $E(w_1, w_2, w_3 \ldots w_n)$.

4) Based on the value of $E$ and the current set of parameters, new parameters are selected and the process is repeated until $<\psi | H_{exact} | \psi>$ is minimized. The Nelder-Mead simplex procedure is a convenient way to accomplish this since the selection of new parameter values depends only on $E$ and the history of its variation with prior choices of parameters.[16]

The result of the optimization procedure is a set of density parameters, $\{c_a, a\}$, for each atom $k$ in the molecule being considered, $\rho_q = \sum_a c_a \rho_a = \sum_a c_a (\frac{a}{\pi})^{3/2} \exp(-a r_q^2)$.

**Acknowledgment**

A collaboration with Dr. Fariba Nazari on metal-metal bonding in transition metal systems is gratefully acknowledged as are helpful discussions with Professor Mike Whangbo.